\pdfoutput=1
\documentclass[11pt,twoside,a4paper,cmspaper,final,collab]{cms-tdr}
\def\svnVersion{44dbce1-D}\def\svnDate{2021/02/26}\def\cmsCernNoTag{CERN-EP-2020-060}\def\cmsCernDate{\today}\def\cmsMessage{"Published in the Journal of High Energy Physics as \href{http://dx.doi.org/10.1007/JHEP05(2021)054}{\doi{10.1007/JHEP05(2021)054}.}"}
\begin{document}\cmsNoteHeader{HIN-18-020}

\hyphenation{back-ground-sub-traction}
\newcommand{\pttrk}{\ensuremath{\pt^{\text{trk}}}\xspace}
\newcommand{\Deltar}{\ensuremath{\Delta r}\xspace}
\newcommand{\Njets}{\ensuremath{N_{\text{jets}}}\xspace}
\newcommand{\Rhodr}{\ensuremath{\rho(\Deltar)}\xspace}
\newlength\cmsTabSkip\setlength{\cmsTabSkip}{1ex}

\cmsNoteHeader{HIN-18-020}

\title{Measurement of $\PQb$ jet shapes in proton-proton collisions at $\sqrt{s} = 5.02\TeV$}

\date{\today}

\abstract{
We present the first study of charged-hadron production associated with jets originating from $\PQb$ quarks in proton-proton collisions at a center-of-mass energy of 5.02\TeV. The data sample used in this study was collected with the CMS detector at the CERN LHC and corresponds to an integrated luminosity of 27.4\pbinv. To characterize the jet substructure, the differential jet shapes, defined as the normalized transverse momentum distribution of charged hadrons as a function of angular distance from the jet axis, are measured for $\PQb$ jets.  In addition to the jet shapes, the per-jet yields of charged particles associated with $\PQb$ jets are also quantified, again as a function of the angular distance with respect to the jet axis. Extracted jet shape and particle yield distributions for $\PQb$ jets are compared with results for inclusive jets, as well as with the predictions from the \PYTHIA and \HERWIGpp event generators. 
} 

\hypersetup{
pdfauthor={CMS Collaboration},
pdftitle={Measurement of b jet shapes in proton-proton collisions at sqrt(s) = 5.02 TeV},
pdfsubject={CMS},
pdfkeywords={CMS, physics, beauty quark, b jets}}

\maketitle

\section{Introduction}

Jets, the collimated showers of particles produced by fragmentation and hadronization of hard-scattered quarks or gluons, are long established experimental probes for studies of quantum chromodynamics (QCD)~\cite{Ali:2010tw}. The internal structure of the jet, defined by the energy, momentum, and spatial distribution of its constituents, is sensitive to the details of the evolution from an initial hard scattering through fragmentation and hadronization into observable hadrons in the final state. The angular distributions of constituent particle yields and jet shapes, studied in this work, are affected by parton fragmentation and hadronization processes. At high transverse momenta ($\pt$) with respect to the beam direction in the core of the jet, the dominant contribution to these distributions is set by the initial branching of the hard scattered parton which is calculable in perturbative QCD (pQCD). 
However, for lower \pt particles and those at larger radial distances from the jet direction, higher order corrections and nonperturbative processes become of major importance. Characterizing the effect of these additional contributions on the internal structure of jets remains challenging for theoretical calculations~\cite{Aaltonen:2008de, Aad:2013fba,Cal:2019hjc}.

In this paper, the internal structure of jets is studied at the charged particle level using the data from proton-proton ($\Pp\Pp$) collisions at a center-of-mass energy of $\sqrt{s}=5.02\TeV$. These data, corresponding to an integrated luminosity of 27.4\pbinv, were collected by the CMS experiment in 2015. For this study, $\PQb$ jets are defined by the presence of at least one $\PQb$ quark, which is inferred from the properties of $\PQb$ hadron decays. A~$\PQb$ jet sample selected via a combined secondary vertex (CSV) discriminator~\cite{Sirunyan:2017ezt}, is composed of jets initiated by a single bottom quark, as well as of a contribution from $\PQb\PAQb$ pairs produced from gluon splitting. Jet-correlated charged particle transverse momentum distributions, referred to as jet shapes, are measured as a function of radial distance $\Deltar = \sqrt{\smash[b]{(\Delta\eta)^{2}+(\Delta\phi)^{2}}}$ from the jet axis. Here $\Delta\eta = \eta^{\text{jet}}-\eta^{\text{trk}}$ and $\Delta\phi = \phi^{\text{jet}}-\phi^{\text{trk}}$ are the pseudorapidity and azimuthal differences between the jet axis and a given charged particle, respectively. To extend the jet shape measurements further into the region where nonperturbative effects dominate, we use a jet-track correlation technique~\cite{HIN_2014_016, Sirunyan:2018jqr}. This method has been shown to reliably subtract the part of the event unrelated to the hard scattering (the underlying event), as well as the contribution of additional pp interactions in the same or nearby bunch crossings (pileup). We study the \pt-differential distributions of jet shapes and particle yields for $\PQb$ jets. By comparing these measurements with the results for inclusive jets and with \HERWIGpp~\cite{Bahr:2008pv} and \PYTHIA~\cite{bib_pythia6_4,bib_pythia8_1} simulations for the $\PQb$ jet and inclusive jet shapes at large angles from the jet axis, this study provides new constraints on pQCD calculations, as well as on the nonperturbative contribution to jet shapes. This measurement also constitutes a baseline for future measurement of the same observable at the same per-nucleon center-of-mass energy in PbPb collisions, which will probe the parton flavor dependence of the interaction of jets with the quark-gluon plasma~\cite{Karsch:1995sy} that is created in high energy heavy-ion collisions. 

\section{The CMS detector}
\label {sec:Detector}
The central feature of the CMS apparatus is a superconducting solenoid of 6\unit{m} internal diameter, providing a magnetic field of 3.8\unit{T}. Within the solenoid volume are a silicon pixel and strip tracker, a lead tungstate crystal electromagnetic calorimeter (ECAL), and a brass and scintillator hadron calorimeter (HCAL), each composed of barrel and endcap sections.  Two forward hadron (HF) steel and quartz-fiber calorimeters complement the barrel and endcap detectors, extending the calorimeter from the range $\abs{\eta}<3.0$ to $\abs{\eta}<5.2$. Events of interest are selected using a two-tiered trigger system~\cite{CMS_run1_trigger}.

In the region $\abs{\eta}<1.74$, the HCAL cells have widths of 0.087 in both pseudorapidity $\eta$ and azimuth $\phi$. Within the central barrel region of $\abs{\eta}<1.48$, the HCAL cells map onto $5\times5$ ECAL crystal arrays to form calorimeter towers projecting radially outwards from the nominal interaction point. Within each tower, the energy deposits in ECAL and HCAL cells are summed to define the calorimeter tower energies, which are subsequently used in the particle flow algorithm to reconstruct the jet energies and directions~\cite{Chatrchyan:2011ds}. In this work, jets are reconstructed within the $\eta$ range of $\abs{\eta}<1.6$.

The silicon tracker measures charged particles within $\abs{\eta}<2.5$. It consists of 1440 silicon pixel and 15\,148 silicon strip detector modules. For nonisolated particles with $1<\pt<10\GeV$ in the barrel region, the track resolutions are typically 1.5\% in \pt and 25--90 (45--150)\mum in the impact parameter direction transverse (longitudinal) to the colliding beams~\cite{Chatrchyan:2014fea}.
A detailed description of the CMS detector, together with a definition of the coordinate system used and the relevant kinematic variables, can be found in Ref.~\cite{Chatrchyan:2008zzk}.

\section{Event selection and simulated event samples}
\label{sec:data_simulation}

The data used in this analysis were taken in a special low-luminosity running period in which there were reduced levels of pileup (approximately 1.5 events per bunch crossing assuming a total inelastic cross section of 65\mbinv~\cite{CMS-PAS-LUM-16-001}). The jet samples are collected with a calorimeter-based trigger that uses the anti-\kt jet clustering algorithm with a distance parameter of $R=0.4$~\cite{bib_antikt}. This trigger requires events to contain at least one jet with $\pt>80\GeV$, and is fully efficient for events containing jets with reconstructed $\pt>90\GeV$. The data selected by this trigger are referred to as "jet-triggered" and are used to study the jet-related particle yields and for data-driven estimation of acceptance effects via an event mixing technique as described in Section~\ref{sec:me_corr}.  To reduce contamination from non-collision events, such as calorimeter noise and beam-gas collisions, vertex and noise reduction selections are applied as described in Refs.~\cite{Chatrchyan:2011sx,Chatrchyan:2012nia}.  These selections include a requirement for events to contain at least $3\GeV$ of energy in one of the calorimeter towers in the HF on each side of the interaction point, and to have a primary vertex (PV) with at least two tracks which are consistent with originating from the same vertex within 15\unit{cm} of the center of the nominal interaction region along the beam axis ($\abs{v_{z}}<15\unit{cm}$).

Monte Carlo (MC) simulated event samples are used to evaluate the performance of the event reconstruction, particularly the track reconstruction efficiency, and the jet energy response and resolution. The MC samples of two different \PYTHIA tunes (version 6.424 with the Z2 tune~\cite{bib:PYTHIA_Z2_and_HERWIG_EE5C} and version 8.230 with the CP5 tune~\cite{bib:PYTHIA8_tune_CP5}) were used to simulate the hard scattering, the parton showering, and the hadronization of the partons. A sample of \PQb jets in MC simulations is obtained from the inclusive simulated QCD jet sample by selecting the jets that are matching to a generator-level $\PQb$ quark within a cone of radius $\Delta R=0.3$~\cite{Chatrchyan:2012jua}. The jets from gluon splitting to $\PQb\PAQb$ are considered as $\PQb$ jets based on this flavour definition. The \GEANTfour (10.02p02)~\cite{bib_geant} toolkit is used to simulate the CMS detector response. An additional reweighting procedure is performed to match the simulated $v_{z}$ distribution to that observed in data. Another QCD jet MC sample is generated using the \HERWIGpp 2.7.1 with the EE5C tune~\cite{bib:PYTHIA_Z2_and_HERWIG_EE5C} and is also used as a theoretical reference.

\section{Jet and track reconstruction}
\label{jet_reco}

Jets are reconstructed offline from the particle-flow (PF) candidates~\cite{Sirunyan:2017ulk}, clustered using the anti-\kt algorithm~\cite{bib_antikt, Cacciari:2011ma} with a distance parameter of $R=0.4$. The PF candidates are reconstructed by the PF algorithm, which aims to reconstruct and identify each particle in an event, with an optimized combination of information from the various elements of the CMS detector. Simulation-derived corrections have been applied to the reconstructed jets to correct the measured energy distortion arising from the limited detector resolution, to the particle level~\cite{Chatrchyan:2011ds,Khachatryan:2016kdb}. Jets with $\pt > 120\GeV$ and $\abs{\eta}<1.6$ are selected to be consistent with a previous study~\cite{Sirunyan:2018jqr}.

A widely used type of the jet axis, the anti-\kt \textit{E}-Scheme jet axis, is calculated by merging all the jet candidates, as well as input particles to the jet clustering by simply adding the four-momenta during the clustering procedure in the anti-\kt algorithm~\cite{Blazey:2000qt}. However, the jet axis for this work is re-calculated by the winner-takes-all recombination scheme~\cite{Bertolini:2013iqa,Larkoski:2014uqa}, which is applied to the constituents found by the nominal anti-\kt \textit{E}-Scheme algorithm for this jet.  

The $\PQb$ jet candidates for this work are selected by the CSV discriminator~\cite{Chatrchyan:2012jua, Chatrchyan:2014fea}. The CSV discriminator is a multivariate classifier that makes use of information about reconstructed secondary vertices (SV) as well as the impact parameters of the associated tracks with respect to the primary vertex, to discriminate $\PQb$ jets from charm-flavor and light-flavor jets. The working point selected for this analysis leads to a 65\% $\PQb$ jet selection efficiency and 69\% purity (the $\PQb$ jet fraction of all jets that passed the CSV selection criteria) from the multijet sample (referring to the background of charm jets and light jets). Possible differences in the purity between data and MC are assessed using a negative-tag technique~\cite{Sirunyan:2017ezt}. This technique selects non-$\PQb$ jets using the same variables and techniques as the standard CSV algorithm both in data and the simulation to extract a scale factor, which indicates the data-to-MC difference. A correction for a bias resulting from the discriminator is discussed in Section~\ref{sec:me_corr}.

In both data and simulation, charged particles are reconstructed using an iterative tracking method~\cite{Chatrchyan:2014fea} based on the hit information from both the pixel and silicon strip subdetectors, permitting the reconstruction of charged particles within $\abs{\eta}<2.4$. The tracking efficiency ranges from approximately 90\% at $\pt=1\GeV$ to no less than 90\% for $\pt>10\GeV$. Tracks with $\pt>1.0\GeV$ and $\abs{\eta}<2.4$ are used in this study. 

\section{Jet-track angular correlations}
\label{sec:me_corr}
To study the distributions of the charged particles associated with jets, a two-dimensional (2D) array of the $\Delta\eta$ and $\Delta\phi$ values of the tracks relative to the jet axis were produced. This is computed for six bins of \pttrk bounded by the values 1, 2, 3, 4, 8, 12, and 300\GeV. Each of these 2D correlations is normalized by $\Njets$, the number of jets in the sample. This procedure, the same one as used in Ref.~\cite{Aad:2014wha}, creates a per-jet averaged $\Delta\eta$-$\Delta\phi$ distribution of raw charged particle densities for each \pttrk:  
\begin{equation}
RS(\Delta\eta, \Delta\phi)=\frac{1}{\Njets}\frac{{\rd}^2N^{\text{same}}}{{\rd}\Delta\eta {\rd}\Delta\phi},
\end{equation}
where $N^{\text{same}}$ represents the yield of jet-track pairs from the same event. For the jet shape measurements, the 2D correlations are weighted by \pttrk on a per-track basis, producing a per-jet averaged $\Delta\eta$-$\Delta\phi$ distribution of \pttrk with respect to the jet axis direction.

An event mixing method~\cite{Sirunyan:2018jqr} is applied following the construction of the raw 2D correlations $RS(\Delta\eta, \Delta\phi)$ to account for the shape of single inclusive jet and track distributions and the effects of the detector acceptance for tracks. For this correction, a mixed-event pair distribution $ME(\Delta\eta, \Delta\phi)$ is constructed by using the jets from one event and the tracks from a different event, matched in the vertex position along the beam axis in 1 cm bins: 
\begin{equation}
ME(\Delta\eta, \Delta\phi)=\frac{1}{\Njets}\frac{\rd^2N^{\text{mix}}}{\rd\Delta\eta \rd\Delta\phi},
\end{equation}
where $N^{\text{mix}}$ represents the number of jet-track pairs from the mixed-event.

The per-jet associated yield is corrected for the jet-track pair efficiency via the following relation:
\begin{equation}
S(\Delta\eta,\Delta\phi)=\frac{1}{\Njets}\frac{\rd^2N}{\rd\Delta\eta {\rd}\Delta\phi}=\frac{ME(0, 0)}{ME(\Delta\eta, \Delta\phi)} RS(\Delta\eta, \Delta\phi).
\label{eq:acceptance_correction}
\end{equation}
The ratio $ME(0, 0)/ME(\Delta\eta, \Delta\phi)$ is the normalized correction factor and $ME(0,0)$ is the mixed event yield for jet-track pairs that are approximately collinear and hence have the maximum pair acceptance. 

The signal of $\PQb$-tagged jets $S_{\text{tag}}(\Delta\eta, \Delta\phi)$ is then corrected for residual light-flavor jet contamination. We use an approach partially relying on data for the decontamination procedure, expressed via the following equation:
\begin{equation}
S_{\text{decont}}(\Delta\eta, \Delta\phi) = \frac{S_{\text{tag}}(\Delta\eta, \Delta\phi)-(1-c_\text{purity})S_{\text{mistagged}}(\Delta\eta, \Delta\phi)}{c_\text{purity}},
\end{equation}\label{eq:decont}
where the $S_\text{decont}(\Delta\eta, \Delta\phi)$ and $S_{\text{mistagged}}(\Delta\eta, \Delta\phi)$ are signals of the (decontaminated) $\PQb$ jets and the mistagged light-flavor jets, respectively. The $S_{\text{mistagged}}(\Delta\eta, \Delta\phi)$ is approximated by the inclusive jet-track correlation signal $S_{\text{inclusive}}(\Delta\eta, \Delta\phi)$ from the data, with a modification for simulating the jet multiplicity bias that is discussed later in this section. The purity $c_\text{purity}$ is defined as the ratio of the number of tagged true jets to the number of jets tagged by the CSV discriminator in simulations.

The resulting decontaminated signal $S_{\text{decont}}(\Delta\eta, \Delta\phi)$ has the residual underlying event contribution and uncorrelated backgrounds from tracks unrelated to selected jets. These backgrounds are then removed in a data-driven manner by using the measured charged-particle yields far from the jet axis in a large-$\Delta\eta$ region. The $\Delta\phi$ distribution averaged over $1.5<\abs{\Delta\eta}<2.5$ is used to estimate the $\Delta\phi$ dependence of the background contribution to the correlations over the entire $\abs{\Delta\eta}<4.0$ region and is subtracted from the signal $S_{\text{decont}}(\Delta\eta, \Delta\phi)$. After that, the background-subtracted signal pair distribution $S(\Deltar)$, as a function of radius $\Deltar$ is obtained from the integration of the corrected signal pair $S(\Delta\eta,\Delta\phi)$, over a ring area with radius $\Deltar$.

The discriminator used for $\PQb$ tagging relies on the properties of the SVs associated with the jet as input, therefore biasing the jet selection towards jets with a better SV or tracking resolution. This bias, though slight, is present in distributions for both true $\PQb$ jets selected by the tagger, and in the mistagged light-flavor jets contaminating the sample.  We calculate corrections for the tagging bias as a function of $\Deltar$ from MC simulation by constructing the following per-jet normalized ratios of radial distributions:
\begin{equation}
\begin{aligned}
B_{\text{mis}}(\Deltar) &= S_{\text{inclusive}}(\Deltar)/S_{\text{mistagged}}(\Deltar),
\\
B_{\PQb}(\Deltar) &= S_{\operatorname{all-\PQb}}(\Deltar)/S_{\operatorname{tagged-\PQb}}(\Deltar),
\label{eq:mis_tag_bias}
\end{aligned}
\end{equation}
where $S_{\text{mistagged}}(\Deltar)$, $S_{\text{inclusive}}(\Deltar)$, $S_{\operatorname{all-\PQb}}(\Deltar)$, and $S_{\operatorname{tagged-\PQb}}(\Deltar)$ represent the signal of tracks correlated with the mistagged jets, inclusive light-flavor jets, and $\PQb$~jets, and the tagged $\PQb$~jets, respectively. This bin-by-bin correction is applied to the background-subtracted signal $S_{\text{decont}}$ to remove the tagging bias.

Finally, simulation-based corrections are applied to account for the jet axis resolution, tracking reconstruction efficiency, and the bias in the charged particle yield and jet shapes that comes from the $\PQb$-tagging discriminator. A large fraction of tracks associated with $\PQb$ jets originate from an SV and have a slightly different reconstruction efficiency from that of tracks originating from a PV. Therefore, we derive the efficiency corrections as a function of track \pt and radial distance $\Deltar$ from the MC $\PQb$ jet simulation by taking the ratio of correlated signals built with reconstructed tracks over those with generated tracks. This bin-by-bin correction has been applied to the signal data distributions obtained in the previous step accordingly. All of these procedures correct the data to a particle level which can be compared with theoretical calculations directly.

The fully corrected 2D correlations are integrated over annular rings in the $\Delta\eta$-$\Delta\phi$ plane (as illustrated in~\cite{Chatrchyan:2012mec}) to study distributions of charged-particle yields $Y(\Deltar)$:
\begin{equation}
Y(\Deltar)= \frac{1}{\Njets} \frac{{\rd}^2N_{\text{trk}}}{\rd \Deltar {\rd}\pttrk}
\end{equation}
with respect to the jet axis as a function of $\Deltar$ for $\PQb$ and inclusive-jet samples and, where $N_{\text{trk}}$ is the number of the charged particles from jets. The jet shape distributions \Rhodr, defined as: 
\begin{equation}
\Rhodr = \frac{1}{\delta r} \frac{\large{\Sigma_{\text{jets}}} \Sigma_{\text{trk} \in ( \Deltar_{\mathrm{a}}, \Deltar_{\mathrm{b}} )} \pttrk}{\large{\Sigma_{\text{jets}}} \Sigma_{\text{trk}} \pttrk}, 
\end{equation}
where $\Deltar_{\mathrm{a}}$ and $\Deltar_{\mathrm{b}}$ define the annular edges of $\Deltar$, $\delta r = \Deltar_{\mathrm{b}} - \Deltar_{\mathrm{a}}$, and $\pttrk$ stands for the $\pt$ of the charged particles, are also examined. 

\section{Systematic uncertainties}
\label{sec:systematics}
A number of sources of systematic uncertainties are considered, including the tracking efficiency, tagging bias corrections, decontamination procedure, jet reconstruction, acceptance corrections, and background subtraction. The systematic uncertainties are summarized in Table~\ref{tab:sys}, and the evaluation of each source of uncertainty is discussed below. 

The tracking reconstruction efficiencies for $\PQb$~jet and inclusive jet tracks have been compared to account for the uncertainty in reconstruction efficiency for displaced tracks, and a maximum difference of about 4\% was observed. The full magnitude of the observed difference is assigned as a conservative estimation to cover the MC-based tracking reconstruction uncertainty.  To study possible differences in track reconstruction between data and simulation, a study of \PD meson decays was used~\cite{CMS-PAS-TRK-10-002}. The \PD meson branching fraction ratio of 3-prong to 5-prong decays was calculated in data with MC-based efficiency corrections and compared with the world-average value~\cite{PDG2018}. The observed difference is used to derive a 4\%  systematic uncertainty for this source. For the full tracking-related uncertainty these two errors were added in quadrature.

The uncertainty for correcting the bias induced by the CSV discriminator is dominated by the uncertainties in the contributions from gluon-splitting and primary $\PQb$ quarks to the $\PQb$~jet sample. Jets originating from different mechanisms of $\PQb$-quark production (i.e. flavor creation, flavor excitation, and gluon splitting) can be studied individually in \PYTHIA simulations. We note that the fraction of $\PQb$ jets from the gluon splitting in \PYTHIA simulation is less than that indicated by data. The corresponding systematic uncertainty has been evaluated by varying this fraction by 20\% (as estimated in Refs.~\cite{Sirunyan:2017ezt,Sirunyan:2018jju}), and the observed 5\% difference in the correction from this variation is propagated as an uncertainty.

The decontamination procedure is affected by the uncertainties in the purity estimation. Using the negative tagging method (described in Section~\ref{jet_reco}) we have derived the data-to-simulation scale factor, which amounted to about 7\% difference in estimated contamination levels. We evaluate the related systematic uncertainty by comparing results obtained with and without the derived scale factor;  less than 5\% variation is observed in the correlation results. This 5\% maximum variation is taken as a systematic uncertainty for the decontamination.

The overall jet energy scale (JES) is sensitive to the relative fraction of quark and gluon jets in the sample. The energy scale uncertainty is found to be 2\% for jets in the study in Ref.~\cite{Khachatryan:2016kdb}. Therefore, we varied the energy threshold of selected jets by this amount in both directions and saw no statistically significant changes in the measured jet shapes. This is not unexpected since the in-jet multiplicity and the jet fragmentation function change slowly with the jet \pt. We also investigated the effects of a more conservative 5\% jet energy scale uncertainty by varying the energy threshold of selected jets by 5\% in both directions and repeating the analysis to uncover any possible differences with respect to the nominal result. The resulting variations in the correlated track yields are found to be below 2\%. Thus, we assigned a 2\% uncertainty for this source. The jet energy resolution (JER) data-to-MC difference is about 15\% based on the $\gamma$+jet studies~\cite{mpt}. The corresponding uncertainty in the reported measurements was evaluated in a data-driven way by smearing the reconstructed jet \pt by 15\% and repeating the study. The resulting variation in correlation distributions was found to be below  3.5\%. In total, a systematic uncertainty of 4\% is assigned for the JER- and JES-related effects.

The uncertainties from the mixed-event acceptance correction are estimated by looking for an asymmetry of the sideband regions, which is defined by the difference of the sideband value between the positive and negative $\Delta\eta$.
Additionally, the sideband regions ($1.5<\abs{\Delta\eta}<2.5$) that are far away from the jet axis are expected to have no short-range correlation contributions and, thus, to be independent of $\Delta\eta$. Any deviations from this expectation and the measured asymmetry are used to quantify the related systematic uncertainty, which was found to be between 1 and 2\%. 

Uncertainties associated with the background subtraction are evaluated by considering the average point-to-point difference between two sideband regions ($1.5 < \abs{\Delta\eta} < 2.0$ and $2.0 < \abs{\Delta\eta} < 2.5$) following the background subtraction. The background subtraction uncertainty is found to be roughly 3\% for the lowest $\pttrk$ bin, where the signal-to-background ratio is the lowest, and decreases to negligible levels as functions of $\pttrk$.

These systematic uncertainties are treated as uncorrelated, and the total systematic uncertainty is calculated by adding the individual sources in quadrature.

\begin{table}[htbp]
\centering
\topcaption{Systematic uncertainties in percentage for the measurements of the jet-track correlations. Where an uncertainty range is given, the upper edge of the range corresponds to the bin with the smallest \pttrk values. The sources from the decontamination and tagging bias are exclusive for $\PQb$ jets.}
\begin{tabular}{lcc}
\hline
Source of systematic uncertainty& $\PQb$ jets & Inclusive jets\\
\hline
Tracking efficiency           		  &5.7  &4.1\\
Tagging bias corrections                 &5.0 &\NA\\
Decontamination procedure              &5.0 &\NA\\
Residual JES and JER corrections       &4.0  &4.0\\
Pair-acceptance corrections              &1.0--2.0 &1.0--2.0\\
Background subtraction                   &0--3.0 &0--3.0\\[\cmsTabSkip]
Total                                    &10.0--10.5 &5.8--6.8\\
\hline
\end{tabular}
\label{tab:sys}
\end{table}

\section{Results}

\begin{figure}[hb!]
     \centering
	\includegraphics[width=1\textwidth]{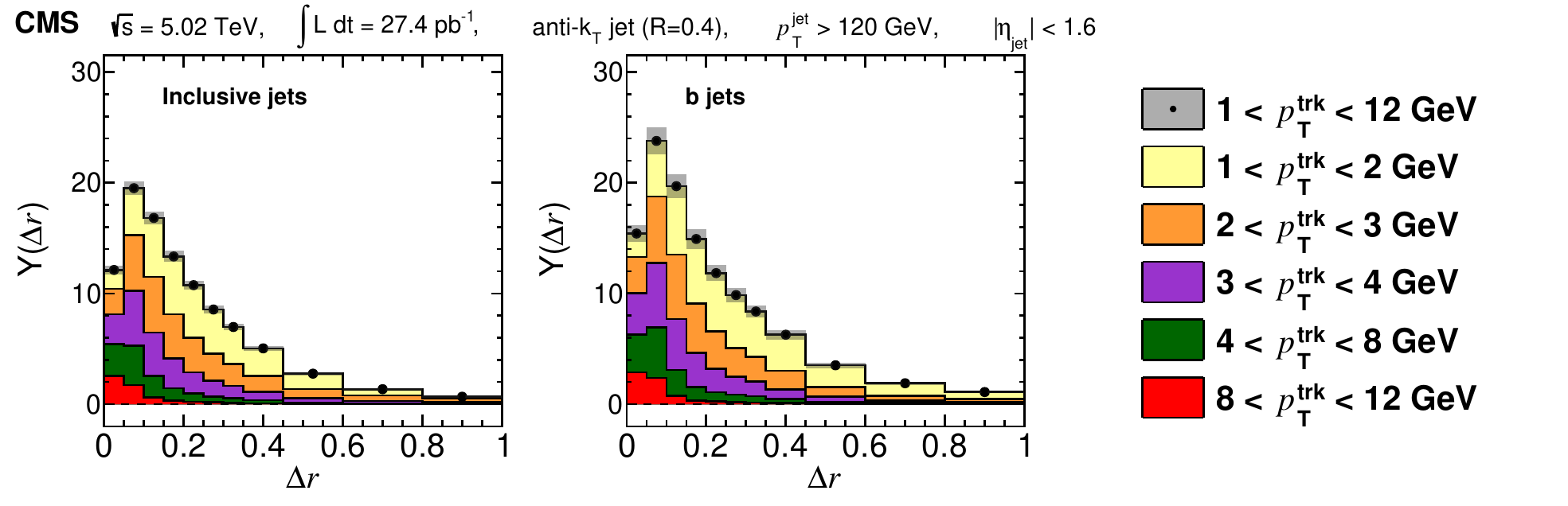}
         \caption{Charged particle yield distributions $Y(\Deltar)$ of inclusive (left) and $\PQb$~(right) jets with $\pt > 120\GeV$ as functions of $\Deltar$ are presented differentially for $\pttrk$ bins. The shadowed boxes represent the systematic uncertainties for $1< \pt^{\text{trk}}<12\GeV$, although they are generally too small to be visible.}
       \label{fig:yield_dr}
 \end{figure}

\begin{figure}[hb!]
     \centering
	\includegraphics[width=1\textwidth]{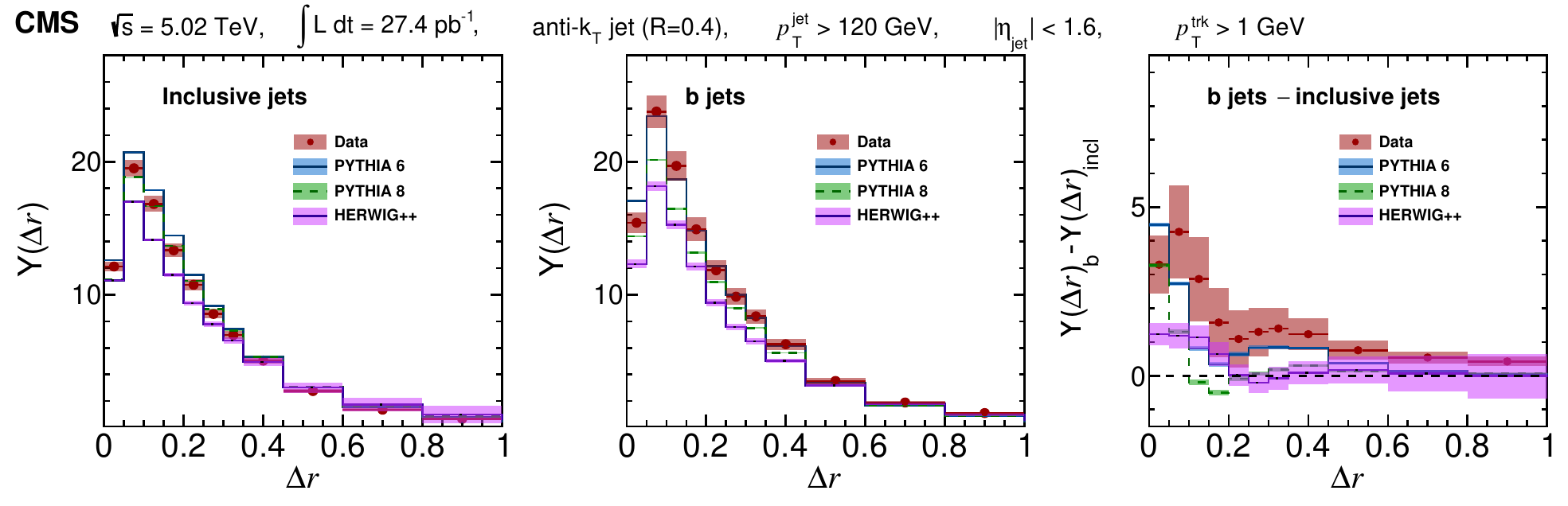}
         \caption{Charged particle yield distributions $Y(\Deltar)$ of inclusive jets (left) and $\PQb$~jets (middle) with $1<\pttrk <12\GeV$ are presented as functions of $\Deltar$. Both types of jets with $\pt > 120\GeV$ and charged particles with $1<\pttrk < 12\GeV$ are used to construct the distributions as functions of $\Deltar$ for data (red),  \PYTHIA 6.426 (blue line), \PYTHIA 8.230 (green dashed line), and \HERWIGpp (purple line) simulations. The right plot shows the particle yield difference of $\PQb$~jets and inclusive jets as functions of $\Deltar$ for $\Pp\Pp$ data, \PYTHIA 6.426 (blue line), \PYTHIA 8.230 (green dashed line) and \HERWIGpp (purple line) simulations. The shadowed boxes represent the systematic uncertainties.}
       \label{fig:yield_dr2}
 \end{figure}

Figure~\ref{fig:yield_dr} presents the charged-particle yields for inclusive and $\PQb$~jets in proton-proton collisions as a function of the radial distance $\Deltar$ from the jet axis. The results are shown with stacked histograms to indicate the intervals in \pttrk, and dots to denote the total summed yields in the region $1 < \pttrk < 12\GeV$. It illustrates that the high-\pt charged particles are mostly distributed around the small $\Deltar$ region while the larger $\Deltar$ region is dominated by the low-\pt charged particles. Figure~\ref{fig:yield_dr2} compares the radial distributions of the total charged-particle yields associated with the inclusive and $\PQb$ jets studied in data and in \PYTHIA simulations. Total uncertainties of the measurement are dominated by the systematic source. Statistical uncertainties of the signal correlations and data-driven mixed-event acceptance correction contribute to the total statistical uncertainties of the data. The total statistical uncertainties are negligible for most data points, except at large $\Delta r$. Statistical uncertainties of the Monte Carlo samples are accounted for in the evaluation of relevant systematic sources and propagated as part of the assigned systematic errors. It is also worth noticing that the systematic uncertainties coming from the event mixing technique are important in the larger $\Delta r$ region. Charged-particle yield distributions for both $\PQb$ and inclusive jets are found to be generally described by \PYTHIA predictions, although \PYTHIA 6.424 shows a better agreement with the data than that found using the \PYTHIA 8.230 predictions. The \HERWIGpp simulation predicts a smaller excess of hadron yields in $\PQb$ jets over inclusive jets compared to what is observed in the data and \PYTHIA simulations. Larger charged-particle yields are observed to be associated with $\PQb$ jets as compared with inclusive jets, particularly in the low-$\Deltar$	region (see Fig.~\ref{fig:yield_dr2}, right). This larger contribution in soft tracks at small radial distance $\Deltar$ implies the presence of different fragmentation patterns and decay kinematics between the $\PQb$ jets and inclusive jets. 

\begin{figure}[hb!]
    \centering
       \includegraphics[width=1\textwidth]{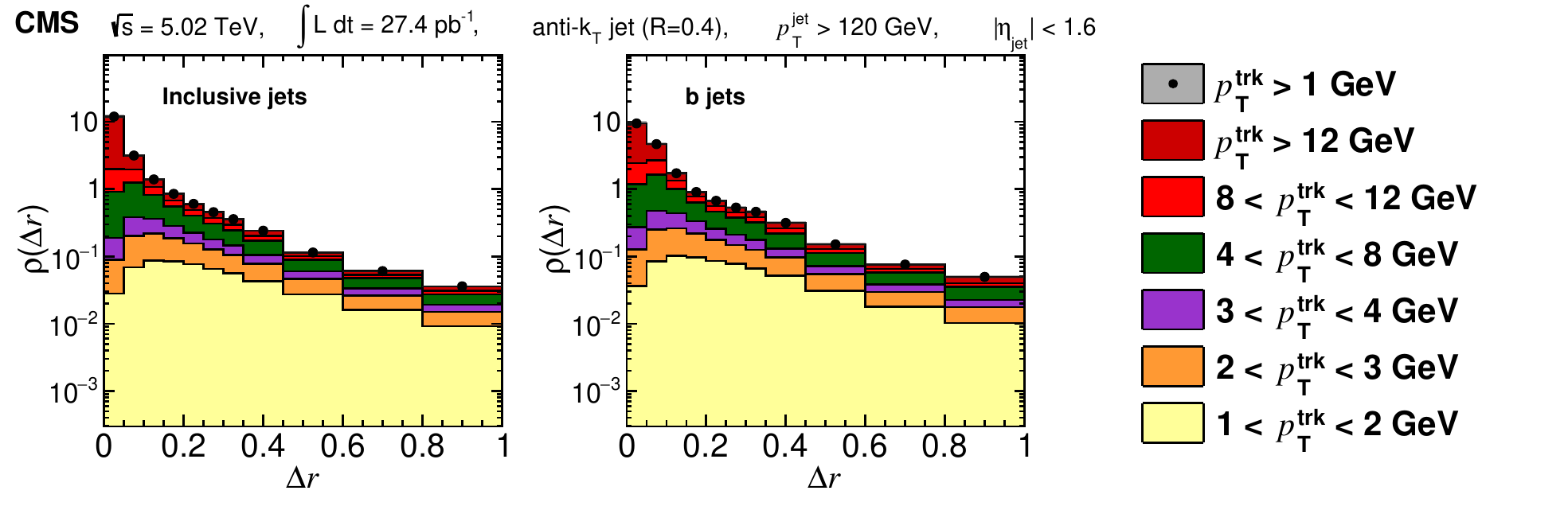}
         \caption{The jet shape distribution $\Rhodr$ of inclusive jets (left) and $\PQb$ jets (right) with $\pt > 120\GeV$ as functions of $\Deltar$ are presented differentially for all $\pttrk$ bins for data. The shadowed boxes represent the systematic uncertainties for $\pt^{\text{trk}}> 1\GeV$, although they are generally too small to be visible.}
       \label{fig:profile_dr}
 \end{figure}

\begin{figure}[hb!]
    \centering
       \includegraphics[width=1\textwidth]{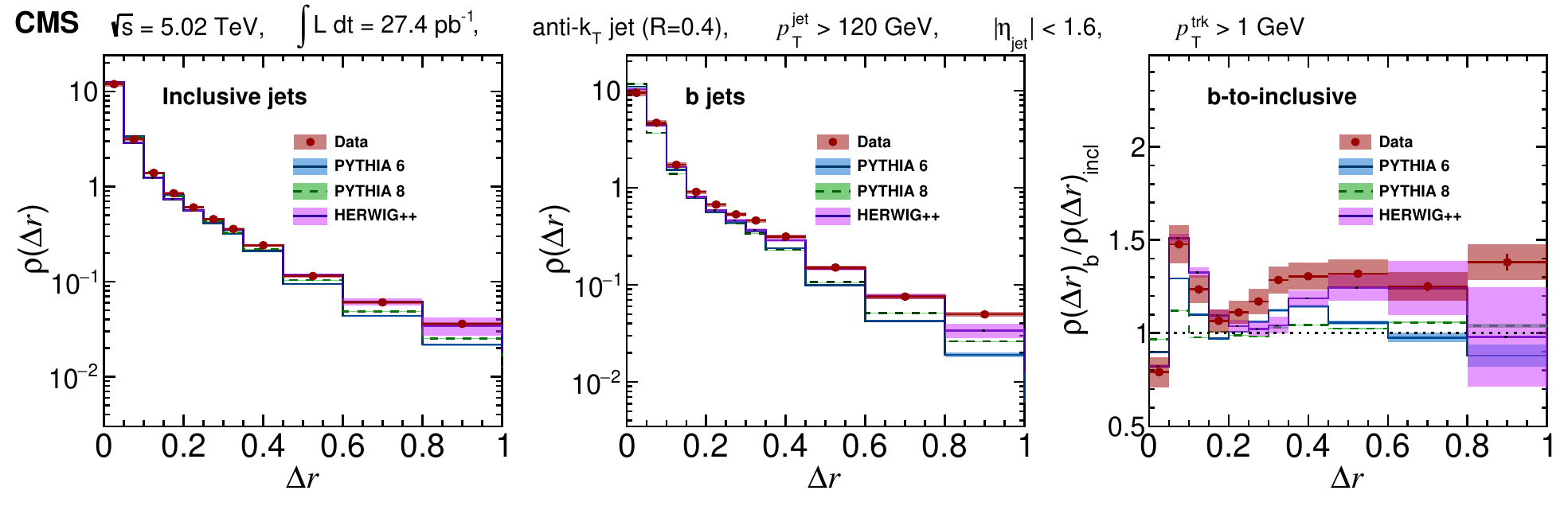}
         \caption{The jet shape distribution $\Rhodr$ of inclusive jets (left) and $\PQb$ jets (middle), both with $\pt > 120\GeV$ and $\pttrk > 1\GeV$ are presented as functions of $\Deltar$ for data(red markers), the \PYTHIA 6.426 (blue line) and the \PYTHIA 8.230 (green dashed line) simulations. The right plot shows the \PQb-to-inclusive jet shape ratio as functions of $\Deltar$ for data, \PYTHIA6 (blue line) and \PYTHIA 8.230 (green dashed line) simulations. The shadowed boxes represent the systematic uncertainty.}
       \label{fig:profile_dr2}
 \end{figure}

Measurements of the jet shapes $\Rhodr$ are presented in Figs.~\ref{fig:profile_dr} and \ref{fig:profile_dr2}.The left and right panels of Fig.~\ref{fig:profile_dr} show \pt-differential $\Rhodr$ distribution for inclusive and $\PQb$ jets, respectively. The comparison between data and simulations from both \PYTHIA and \HERWIGpp is presented in Fig.~\ref{fig:profile_dr2}. We note that, while small-$\Deltar$ trends are mostly well described by \PYTHIA and \HERWIG simulation for both jet selections, the distributions at larger radial distances are only well-estimated by \HERWIGpp, indicating a shortage of soft radiative contributions. The right panel of Fig.~\ref{fig:profile_dr2} shows the ratio of  $\PQb$ to inclusive jet shapes for data and simulation.

Observed variations in the ratio of jet shapes indicate a shift of transverse momentum from small to large $\Deltar$ for the constituents of the $\PQb$ jets compared to that carried by the particles from inclusive jets. These differences may arise from the dead-cone effect, the suppression of radiation from a charged particles with mass $m_\cPq$ and energy $E_\cPq$ in the region with emission angle $\theta \lesssim m_\cPq/E_\cPq$~\cite{Dokshitzer:1991fc,Dokshitzer:1991fd}, as this phenomenon is expected to be more apparent in $\PQb$ jets than in inclusive jets, which mostly originate from light partons. 
	
\PYTHIA and \HERWIG simulations show very different jet shape predictions at large angular distances, where nonperturbative contributions are likely to dominate. \HERWIGpp simulations have better performance capturing the details of jet shapes for both inclusive and $\PQb$ jets distributions in this region, comparing to \PYTHIA simulations. Additionally, we observe that a higher fraction of transverse momentum is distributed towards the higher radial distances from the center of the jet for the $\PQb$ jets as compared to the inclusive jet sample. 

A similar tendency, albeit insufficient to fully capture this trend, is seen in \PYTHIA simulations. \PYTHIA 8.230 simulations show a slightly better description than that from \PYTHIA 6.426 in the larger $\Deltar$ region. On the other hand, \HERWIGpp 2.7.1 predictions capture this trend well, as illustrated in the right panel of Fig.~\ref{fig:profile_dr2}. 
The observed data to \PYTHIA discrepancy in the \PQb-to-inclusive jet shape ratios at large radii may arise from the difference in the gluon splitting contributions between data and simulation, as mentioned earlier~\cite{Banfi:2007gu}. We note that Monte Carlo studies show that $\PQb$ and $\PAQb$ jets from gluon splitting result in significantly broader jet shapes than those of inclusive jets.

\section{Summary}
\label{sec:Summary}
The first measurements of charged-particle yields and jet shapes for $\PQb$ jets in proton-proton collisions are presented, using data collected with the CMS detector at the LHC at a center-of-mass energy of $\sqrt{s} = 5.02\TeV$. The correlations of charged particles with jets are studied, using the particles with transverse momentum $\pttrk > 1\GeV$ and pseudorapidity $\abs{\eta}<2.4$, and the jets with $\pt > 120\GeV$ and $\abs{\eta}<1.6$. Charged-particle yields associated with jets are presented as functions of the relative angular distance $\Deltar = \sqrt{\smash[b]{(\Delta\eta)^{2}+(\Delta\phi)^{2}}}$ from the jet axis. In these studies, a large number of associated charged particles at low $\Deltar$ are found for $\PQb$~jets compared to those for inclusive jets, which are produced predominantly by gluons and light flavor quarks. The trends observed in $\Pp\Pp$ data for particle yield distributions associated with both types of jets are reproduced by \PYTHIA calculations (in versions 6.426 and 8.230).

In addition to the charged-particle yields, we examine the jet transverse momentum profile variable $\Rhodr$, defined using the distribution of charged particles in annular rings around the jet axis, with each particle weighted by its \pttrk value. The measured shapes of $\PQb$ jets are broader than those of inclusive jets. The shapes for both types of jets are reproduced by \HERWIG and \PYTHIA calculation in the small $\Deltar$ region, with \HERWIGpp 2.7.1 giving a better agreement. Moreover, measured transverse momenta distributions at larger $\Deltar$ are consistent with the \HERWIG simulations for $\PQb$ and inclusive jets, with at most 1.2 $\sigma$ data-to-simulation differences observed for $\PQb$ jets. However, this trend is generally underestimated by \PYTHIA simulations.

This result provides new constraints on perturbative quantum chromodynamics calculations for flavor dependence in parton fragmentation and gluon radiation, as well as the relative contributions of different processes to $\PQb$ quark production. These measurements are also expected to offer an important reference for future studies of flavor dependence for parton interactions with the quark-gluon plasma formed in relativistic heavy-ion collisions.

\begin{acknowledgments}
  We congratulate our colleagues in the CERN accelerator departments for the excellent performance of the LHC and thank the technical and administrative staffs at CERN and at other CMS institutes for their contributions to the success of the CMS effort. In addition, we gratefully acknowledge the computing centers and personnel of the Worldwide LHC Computing Grid for delivering so effectively the computing infrastructure essential to our analyses. Finally, we acknowledge the enduring support for the construction and operation of the LHC and the CMS detector provided by the following funding agencies: BMBWF and FWF (Austria); FNRS and FWO (Belgium); CNPq, CAPES, FAPERJ, FAPERGS, and FAPESP (Brazil); MES (Bulgaria); CERN; CAS, MoST, and NSFC (China); COLCIENCIAS (Colombia); MSES and CSF (Croatia); RPF (Cyprus); SENESCYT (Ecuador); MoER, ERC IUT, PUT and ERDF (Estonia); Academy of Finland, MEC, and HIP (Finland); CEA and CNRS/IN2P3 (France); BMBF, DFG, and HGF (Germany); GSRT (Greece); NKFIA (Hungary); DAE and DST (India); IPM (Iran); SFI (Ireland); INFN (Italy); MSIP and NRF (Republic of Korea); MES (Latvia); LAS (Lithuania); MOE and UM (Malaysia); BUAP, CINVESTAV, CONACYT, LNS, SEP, and UASLP-FAI (Mexico); MOS (Montenegro); MBIE (New Zealand); PAEC (Pakistan); MSHE and NSC (Poland); FCT (Portugal); JINR (Dubna); MON, RosAtom, RAS, RFBR, and NRC KI (Russia); MESTD (Serbia); SEIDI, CPAN, PCTI, and FEDER (Spain); MOSTR (Sri Lanka); Swiss Funding Agencies (Switzerland); MST (Taipei); ThEPCenter, IPST, STAR, and NSTDA (Thailand); TUBITAK and TAEK (Turkey); NASU (Ukraine); STFC (United Kingdom); DOE and NSF (USA). 
  
  \hyphenation{Rachada-pisek} Individuals have received support from the Marie-Curie program and the European Research Council and Horizon 2020 Grant, contract Nos.\ 675440, 752730, and 765710 (European Union); the Leventis Foundation; the A.P.\ Sloan Foundation; the Alexander von Humboldt Foundation; the Belgian Federal Science Policy Office; the Fonds pour la Formation \`a la Recherche dans l'Industrie et dans l'Agriculture (FRIA-Belgium); the Agentschap voor Innovatie door Wetenschap en Technologie (IWT-Belgium); the F.R.S.-FNRS and FWO (Belgium) under the ``Excellence of Science -- EOS" -- be.h project n.\ 30820817; the Beijing Municipal Science \& Technology Commission, No. Z191100007219010; the Ministry of Education, Youth and Sports (MEYS) of the Czech Republic; the Deutsche Forschungsgemeinschaft (DFG) under Germany's Excellence Strategy -- EXC 2121 ``Quantum Universe" -- 390833306; the Lend\"ulet (``Momentum") Program and the J\'anos Bolyai Research Scholarship of the Hungarian Academy of Sciences, the New National Excellence Program \'UNKP, the NKFIA research grants 123842, 123959, 124845, 124850, 125105, 128713, 128786, and 129058 (Hungary); the Council of Science and Industrial Research, India; the HOMING PLUS program of the Foundation for Polish Science, cofinanced from European Union, Regional Development Fund, the Mobility Plus program of the Ministry of Science and Higher Education, the National Science Center (Poland), contracts Harmonia 2014/14/M/ST2/00428, Opus 2014/13/B/ST2/02543, 2014/15/B/ST2/03998, and 2015/19/B/ST2/02861, Sonata-bis 2012/07/E/ST2/01406; the National Priorities Research Program by Qatar National Research Fund; the Ministry of Science and Higher Education, project no. 02.a03.21.0005 (Russia); the Tomsk Polytechnic University Competitiveness Enhancement Program and ``Nauka" Project FSWW-2020-0008 (Russia); the Programa Estatal de Fomento de la Investigaci{\'o}n Cient{\'i}fica y T{\'e}cnica de Excelencia Mar\'{\i}a de Maeztu, grant MDM-2015-0509 and the Programa Severo Ochoa del Principado de Asturias; the Thalis and Aristeia programs cofinanced by EU-ESF and the Greek NSRF; the Rachadapisek Sompot Fund for Postdoctoral Fellowship, Chulalongkorn University and the Chulalongkorn Academic into Its 2nd Century Project Advancement Project (Thailand); the Kavli Foundation; the Nvidia Corporation; the SuperMicro Corporation; the Welch Foundation, contract C-1845; and the Weston Havens Foundation (USA). 
\end{acknowledgments}

\bibliography{auto_generated}

\cleardoublepage \appendix\section{The CMS Collaboration \label{app:collab}}\begin{sloppypar}\hyphenpenalty=5000\widowpenalty=500\clubpenalty=5000\vskip\cmsinstskip
\textbf{Yerevan Physics Institute, Yerevan, Armenia}\\*[0pt]
A.M.~Sirunyan$^{\textrm{\dag}}$, A.~Tumasyan
\vskip\cmsinstskip
\textbf{Institut f\"{u}r Hochenergiephysik, Wien, Austria}\\*[0pt]
W.~Adam, F.~Ambrogi, T.~Bergauer, M.~Dragicevic, J.~Er\"{o}, A.~Escalante~Del~Valle, M.~Flechl, R.~Fr\"{u}hwirth\cmsAuthorMark{1}, M.~Jeitler\cmsAuthorMark{1}, N.~Krammer, I.~Kr\"{a}tschmer, D.~Liko, T.~Madlener, I.~Mikulec, N.~Rad, J.~Schieck\cmsAuthorMark{1}, R.~Sch\"{o}fbeck, M.~Spanring, W.~Waltenberger, C.-E.~Wulz\cmsAuthorMark{1}, M.~Zarucki
% ...
\vskip\cmsinstskip
\dag: Deceased\\
1:  Also at Vienna University of Technology, Vienna, Austria\\
% ...
\end{sloppypar}
\end{document}